\documentstyle[aps]{revtex} 
\newcommand{\el}{{\bf \tilde{E}\/}}
\newcommand{\mg}{{\bf \tilde{B}\/}} 
\newcommand{\elo}{{\bf \tilde{E}_{0}\/}} 
\newcommand{\mgo}{{\bf \tilde{B}_{0}\/}}
\newcommand{\bb}{{\bf p \/}} 
\newcommand{\kk}{{\bf k \/}} 
\begin{document}
\title{Cosmic optical activity from an inhomogeneous Kalb-Ramond field}
\author{Sayan Kar\footnote{Electronic address: {\em
sayan@cts.iitkgp.ernet.in}}$^1$, Parthasarathi
Majumdar\footnote{Electronic address: {\em partha@imsc.ernet.in}}$^2$,
Soumitra SenGupta\footnote{Electronic address: {\em
soumitra@juphys.ernet.in}}$^3$, and Saurabh Sur\footnote{Electronic
address: {\em saurabh@juphys.ernet.in}}$^4$} 
\address{{\rm $^1$}Centre for
Theoretical Studies and Department of Physics, Indian Institute of
Technology, Kharagpur 721302, India} 
\address{{\rm $^2$}Institute of
Mathematical Sciences, Chennai 600113, India} 
\address{{\rm $^{3,4}$}Department of Physics, Jadavpur University,
Calcutta 700 032,
India} 
\maketitle 
\begin{abstract} 

The effects of introducing a harmonic spatial inhomogeneity into the
Kalb-Ramond field, interacting with the Maxwell field according to a
`string-inspired' proposal made in earlier work are investigated. We
examine in particular the effects on the polarization of synchrotron
radiation from cosmologically distant (i.e. of redshift greater than 2)
galaxies, as well as the relation between the electric and magnetic
components of the radiation field. The rotation of the polarization plane
of linearly polarized radiation is seen to acquire an additional
contribution proportional to the square of the frequency of the dual
Kalb-Ramond axion wave, assuming that it is far smaller compared to the
frequency of the radiation field.
\end{abstract}

\section{Introduction}

The behaviour of electromagnetic waves in a curved background spacetime
with torsion and its cosmological consequences, has been an area of some
interest in recent years. This is in view of its prospective implications
for the low energy approximation to string theory. One way to investigate
these implications is by identifying spacetime torsion \cite{pmss}, with
the massless antisymmetric second rank tensor field, i.e., the Kalb-Ramond
(KR) field, existing in most supergravity theories and as such in the
massless sector of the most viable string theories \cite{gsw}. Certain
physically observable phenomena result from the above analysis: a cosmic
optical activity involving the rotation of the plane of polaization of
linearly polarized synchrotron rotation from high redshift galaxies
investigated recently \cite{skpm} being an example. The angle through
which the polarization plane rotates is shown to be proportional, to
leading order in inverse conformal time (which decreases with redshift),
to the rate of change of the KR dual axion field. It is also independent
of the frequency of radiation. This latter aspect is in contrast to the
well-known Faraday rotation, and hence a new phenomenon. 

Another point to note is that the KR field, argued to be responsible for
the effect, has been treated in \cite{skpm} as a perturbation on the
Maxwell equations in a standard Friedmann-Robertson-Walker (FRW)
cosmological background with both matter and radiation domination. Thus,
it is assumed to have a negligible effect on shaping cosmological
background spacetime. One way of thinking about this is to imagine that
the KR axion decouples from the radiation or matter (dust) fluid shaping
cosmic geometry far prior to dust-photon decoupling, leaving behind a
`Cosmic KR Background' which affects incoming radiation from distant
galaxies, albeit rather softly. As the universe expands further, this
effect will no doubt gradually subside. The point made in \cite{skpm} is
that the effect may yet be observable in this epoch. This viewpoint has
received support from \cite{pdpj} where the the effects of a
time-dependent dilaton field are additionally incorporated, while
demonstrating that the earlier findings have a degree of robustness.

However, in these earlier assays, it is assumed that the axion and dilaton
fields are spatially homogeneous, depending only on the conformal time
coordinate $\eta$. This is possibly quite justified considering the
overall homogeneity of the universe over cosmological distance scales.

In this paper, we generalize the scenario in \cite{skpm} by introducing a
spatial inhomogeneity of a particular type into the KR axion field $H$:
the massless Klein-Gordon equation obeyed by the axion quite naturally
leads us to take the spatial dependence in the form of a plane wave
propagating in space. However, taking into consideration the fact that the
effects caused by the KR field should be confined to a very feeble
disturbance on the overall homogeneity of space, we assume the the
frequency of the axionic wave to be far smaller compared to that of the
electromagnetic wave. This modification produces some interesting
features, such as, it alters the mutual orthogonality of the electric and
magnetic field vectors while inflicting a change on the Poynting equation
as well. Furthermore, the inhomogeneity produces an additional rotation of
the plane of polarization of radiation over that found in \cite{skpm}.
This additional contribution turns out to be porportional to the square of
the frequency of the axion field while being independent of the wavelength
of the radiation.
 
The paper is organized as follows. For completeness and as a background,
we briefly recapitulate in Section 2 the basic tenets of \cite{skpm},
which lead to a modified set of Maxwell equations. The harmonic spatial
dependence assumed for the axion field further modifies these equations as
well as the one for the H field itself. Demanding that the
waves retain their forms obtained in \cite{skpm} in the limit $H$ becomes
space-independent, we obtain the equations representing
circularly-polarized states.  Carrying out a standard WKB type procedure,
we solve these equations and calculate the rotation angle of the plane of
polarization of the fields for a flat background spacetime in Section 3
and for a spatially flat spacetime in Section 4, considering separately
the radiation and matter dominated cases therein. We make a few concluding
remarks on our results in Section 5.

\section{Gauge invariant Einstein-Cartan-Maxwell-Kalb-Ramond coupling}
\subsection{Modified field equations}

The action for gauge-invariant Einstein-Cartan-Maxwell-Kalb-Ramond
coupling is taken to be of the form \cite{pmss}:

\begin{equation}
S = \int~ d^{4}x \sqrt{-g} ~\left[~\frac{\tilde{R} (g,T)}{\kappa} -
    \frac{1}{4} F_{\mu \nu} F^{\mu \nu} -
    \frac{1}{2} \tilde{H}_{\mu \nu \lambda} \tilde{H}^{\mu \nu \lambda} +
    \frac{1}{\sqrt{\kappa}} T^{\mu \nu \lambda} \tilde{H}_{\mu \nu 
    \lambda}~\right]
\end{equation}

\noindent
$\tilde{R}(g,T)$ being the scalar curvature for the Einstein-Cartan 
spacetime where the connection contains the torsion tensor 
$T_{\alpha \mu \nu}$ (supposed to be antisymmetric in all its indices)
in addition to the Christoffel term; $\kappa = 16 \pi G$ is the 
coupling constant; and $\tilde{H}_{\mu \nu \lambda}$ is the KR field 
strength three-tensor modified by U(1) Chern-Simons term arising form 
the quantum consistency of an underlying string theory:
\begin{equation}
\tilde{H}_{\mu \nu \lambda} = \partial_{[\mu} B_{\nu \lambda]} +
\frac{1}{3} \sqrt{\kappa} A_{[\mu} F_{\nu \lambda]},
\end{equation}
with, $B_{\nu \lambda}$ being the antisymmetric KR potential which is
considered to be the possible source of torsion. $\tilde{R}$ is related to
the scalar curvature $R$ of purely Riemannian (torsion-free) space-time by 
\begin{equation}
\tilde{R}(g,T) = R(g) + T_{\mu \nu \lambda} T^{\mu \nu \lambda},
\end{equation}

The fact that the augmented KR field strength three tensor plays the role 
of spin angular momentum density (which is the source of torsion \cite{hehl}) 
can be evidenced directly from Eq.(1) where the torsion tensor $T_{\mu \nu 
\lambda}$, being an auxiliary field, obeys the constraint equation
\begin{equation}
T_{\mu \nu \lambda} = \sqrt{\kappa} \tilde{H}_{\mu \nu \lambda}.
\end{equation}
Substituting Eq.(4) in the action (1) and varying the latter with respect to 
$B_{\mu \nu}$ and $A_{\mu}$ respectively, two sets of field equations are 
obtained
\begin{equation}
D_{\mu} \tilde{H}^{\mu \nu \lambda} \equiv 
\frac{1}{\sqrt{- g}} \partial _{\mu}
(\sqrt{- g} \tilde{H}^{\mu \nu \lambda}) = 0 
\end{equation}
and
\begin{equation}
D_{\mu} F^{\mu \nu} = \sqrt{\kappa} \tilde{H}^{\mu \nu \lambda}F_{\lambda \mu}.
\end{equation}

In addition, there is also the Maxwell - Bianchi identity
\begin{equation}
D_{\mu}~ ^{*}F^{\mu \nu} \equiv \frac{1}{\sqrt{- g}} \partial_{\mu}
(\sqrt{- g}~ ^{*}F^{\mu \nu}) = 0.
\end{equation}
Note here that all covariant derivatives are defined with the Christoffel
connection and the Maxwell field strength is the standard 2-form $F=dA$.

Now, expressing the KR field strength three tensor $H_{\mu \nu \lambda} \equiv
\partial_{[\mu} B_{\nu \lambda]}$  as the Hodge-dual to the derivative of the 
spinless pseudoscalar field $H$ (the axion):
\begin{equation}
H_{\mu \nu \lambda} = \epsilon_{\mu \nu \lambda}^{\rho} D_{\rho} H.
\end{equation}
and substituting in Eqs.(6) and (7), the modified generally 
covariant Maxwell's equations are obtained in three-vectorial form 
\begin{eqnarray}
{\bf D\/}\cdot{\bf E\/} &=& 2 \sqrt{\kappa}~{\bf D\/} H\cdot{\bf B\/} \\
D_{0} {\bf E\/} ~-~ {\bf D\/} \times {\bf B\/} &=& 
-~2 \sqrt{\kappa}~[~D_{0}H~{\bf B\/} ~-~ {\bf D\/}H \times {\bf E\/}~] 
~+~ O(\kappa)\\
{\bf D\/}\cdot{\bf B\/} &=& 0 \\
D_{0} {\bf B\/} ~-~ {\bf D\/} \times {\bf E\/} &=& 0 
\end{eqnarray}
where $D_{\mu}$ stands for the covariant derivative. On dropping all the higher 
order terms, as a first approximation, and retaining terms only of the order of 
$\sqrt{\kappa}$, in a spatially flat isotropic FRW background with metric 
\begin{equation}
ds^{2} = R^{2}(\eta) (d\eta^{2} - d{\bf x\/}^{2}),
\end{equation}
the set of equations (9)-(12) take the form
\begin{eqnarray}
\nabla \cdot \el &=& 2 \nabla H \cdot \mg \\
\partial_{\eta} \el ~-~ \nabla \times \mg &=& -2[ ~\partial_{\eta} H \mg
- \nabla H \times \el~] \\
\nabla \cdot \mg &=& 0 \\
\partial_{\eta} \mg ~+~ \nabla \times \el &=& 0 
\end{eqnarray}
where $\eta$ the conformal time coordinate, defined by $d\eta = dt/R(t)$,
$R$ is the cosmological scale factor; and $\el = R^{2}{\bf E\/}$ and 
$\mg = R^{2}{\bf B\/}$. $H$ is redefined by absorbing the $\sqrt{\kappa}$ in it.

It is easy to show from the very form of the KR field strength, viz., 
$H_{\mu \nu \lambda} = \partial_{[\mu}B_{\nu \lambda]}$, that it 
satisfies the Bianchi identity 
\begin{equation}
\epsilon^{\mu \nu \lambda \rho} \partial_{\rho} H_{\mu \nu \lambda} = 0
\end{equation}
which immediately implies that $H$ satisfies the wave equation
\begin{equation} 
D_{\rho} D^{\rho} H = 0 
\end{equation}

In isotropic spatially flat universe this equation reduces to
\begin{equation}
(\partial_{\eta}^{2} ~-~ \nabla^{2}) H ~= ~-~2 \frac{\dot{R}}{R} \dot{H}
\end{equation} 
where $H$ is taken to be a general function of both space and time 
coordinates, and the over-dot implies partial differentiation with 
respect to $\eta$. Assuming a general wave solution for $H$, viz.,
\begin{equation}
H(\eta,{\bf x\/}) ~=~ H_{0}(\eta)~ \cos ~{\bb \cdot {\bf x\/}}
\end{equation}
Eq.(20) enables us to get an equation for $H$
\begin{equation}
\ddot{H} ~+~ 2 \frac{\dot{R}}{R} \dot{H} ~+~ p^{2} H ~=~ 0
\end{equation}

We should point out here that as a consequence of our prior assumption
that the overall homogeneity of the universe over long distance scales is not 
much disturbed by the inclusion of the spatial part in $H$, we are taking 
$p$ to be much less compared to the wave number for the electromagnetic
radiation. 

\subsection{ Modifications in electro-magnetic orthogonality and
the Poynting equation}

{}From the field equations (14) - (17) we derive the wave equations for
the electric and magnetic fields
\begin{eqnarray}
\Box \mg ~\equiv~ (~\partial_{\eta}^{2} ~-~ \nabla^{2}~)~ \mg &=&
2~ \nabla \times (~\dot{H} \mg) ~-~ 2~ \nabla \times (~\nabla H \times \el~) \\
\Box \el ~\equiv~ (~\partial_{\eta}^{2} ~-~ \nabla^{2}~)~ \el &=&
2~ \nabla \times (~\dot{H} \el) ~+~ 2~ \nabla H \times \dot{\el} ~-~ 
\nabla (~\nabla \cdot \el~) ~-~ 2~ \ddot{H} \mg 
\end{eqnarray}
These equations indeed reduce to the pure Maxwell equations in the limit 
$H ~\rightarrow ~0$, or a constant. Treating the axion $H$ as a {\it tiny}
purturbation over the Maxwell equations we argue that the solutions should
have a form not much departing from the usual plane wave structure with the
wave vector $\kk$ perpendicular to both the electric and magnetic vectors.
Moreover, the field equations (14) and (15) enable us to derive 
\begin{equation}
\nabla \dot{H} \cdot \mg = 0
\end{equation}
which, in view of the specific form of $H$, viz., $H_{0}(\eta)~ \cos 
({\bb \cdot {\bf x\/}})$, implies 
\begin{equation}
\bb \cdot \mg = 0
\end{equation}
provided $\dot{H} \ne 0$, which is the general case we are handling.

This orthogonality of $\bb$ and $\mg$ [Eq.(26)] makes it easier to assume, for
simplicity, that $\bb$ can be taken to be orthogonal to $\el$ as well, i.e., 
$\bb$ is either parallel or antiparallel to $\kk$. This is fairly justified, 
as it looks, from the similar wave nature of the electric and magnetic fields, 
at least in the limiting pure Maxwellian case as we are treating the modification 
caused by the KR field as a small purturbation over the Maxwellian behaviour. 
In fact, we may choose to observe the electromagnetic radiation which is travelling
in the direction of propagation of the KR field.

Considering the z-direction to be the propagation direction of the
electromagnetic waves and as such for the axion (following the abovementioned
assumption) we reduce the four-dimesional problem to a two-dimensional one with
$\eta$ and $z$ being the only variables. The field equations now reduce to  
simpler forms
\begin{eqnarray}
\Box \mg &=& 2~ \dot{H} \nabla \times \mg ~+~ 2  \partial_{\eta} 
(H' {\bf \hat{e}_{z}} \times \mg) ~+~ 2~ H'' \el\\
\Box \el &=& 2~ \dot{H} \nabla \times \el ~+~ 2  \partial_{\eta} 
(H' {\bf \hat{e}_{z}} \times \el) ~-~ 2~ \ddot{H} \mg 
\end{eqnarray}
where the over-dot and prime denote respectively the partial
differentiations with respect to $\eta$ and $z$; and ${\bf \hat{e}_{z}}$
is the unit vector along the z-direction. Now, it is easy to show from the 
field equations (15) and (17) that
\begin{equation}
\partial_{\eta}~(\el \cdot \mg)~ = ~\mg \cdot \nabla \times \mg ~-
\el \cdot \nabla \times \el ~-~ 2~\dot{H}~\mg^2 ~+~ 2~\nabla H \cdot
\el \times \mg
\end{equation}

Considering the limiting plane wave behaviour of the solutions of the 
electromagnetic wave equations as $~H ~\rightarrow ~0~$, the magnetic and the
electric fields can be expressed 
\begin{eqnarray}
\mg (\eta,z) &=& \mgo (\eta,z) ~ e^{- i k z} \\
\el (\eta,z) &=& \elo (\eta,z) ~ e^{- i k z}
\end{eqnarray}

When the KR field actually vanishes, $\elo, \mgo ~\equiv~$ constant
vectors~$\times~ e^{i k \eta}$ ~in the plane wave solutions of the pure
Maxwell equations. Eq.(29)  then gives $~\el \cdot \nabla
\times \el ~=~ \mg \cdot \nabla \times \mg ~=~ 0$, i.e., $\partial_{\eta}
(\el \cdot \mg) ~=~ 0$. Moreover, since $\mg$ is in the direction of
$\nabla \times \el$, as is evident from Eq.(17), it follows that $\el
\cdot \mg ~=~ 0$. When the KR field is present, but is only
time-dependent (the case in \cite{skpm}), the vectors $\elo$ and $\mgo$
in the solutions (30) and (31) again depend on time only and Eq.(29)
reduces to
\begin{equation}
\partial_{\eta}~(\elo \cdot \mgo)~ = ~-~ 2~\dot{H}~\mgo^2.
\end{equation}

Clearly, $~\elo \cdot \mgo ~\ne~ 0$,~ which implies that the mutual 
orthogonality of $\el$ and $\mg$ is lost. In the most general case where
the KR field depends on both space and time coordinates, $\elo$ and $\mgo$
are also spacetime-dependent and they satisfy the relation
\begin{equation}
\partial_{\eta}~(\elo \cdot \mgo)~ = ~\mgo \cdot \nabla \times \mgo ~-~
\elo \cdot \nabla \times \elo ~-~ 2~\dot{H}~\mgo^2 ~+~ 2~\nabla H \cdot
\elo \times \mgo
\end{equation}
Here we find, one cannot ascertain conclusively that $\elo \cdot \mgo$ is
manifestly zero. It is, in fact, more justified to think that $\elo \cdot
\mgo$ is essentially non-zero in general, at least, by looking at the
limiting behaviour when the KR field is stripped off the spatially
depending part. The last term in Eq.(33), appearing due to the inclusion
of the spatial dependence in $H$ cannot, in general, compensate for the
term which actually renders $\elo \cdot \mgo$ non-vanishing in Eq.(32), as
the temporal and spatial components of the KR field are completely
separate entities.
  
The Poynting equation in the present scenario can be obtained directly
from the field equations (14) - (17). It is given by
\begin{equation}
\nabla \cdot {\bf S} ~+~ \dot {\omega}_{em} ~=~ - 2 \dot{H}~ \el 
\cdot \mg
\end{equation}
where ~$ {\bf S} ~=~ (\el \times \mg) $~ is the Poynting vector and 
~$ \omega_{em} ~=~ \frac 1 2 (\el^2 ~+~ \mg^2) $~ is the electromagnetic
energy density. The distinction of this equation over the Poynting
equation in the pure Maxwellian case is the presence of the term 
on the right hand side which is, in general, non-zero for reasons
discussed above. In fact, this term vanishes in the limit ~$\dot{H} ~
\rightarrow~ 0$,~i.e., when $H$ becomes purely space-dependent. This is quite
obvious since spatial inhomogeneity alone cannot bring in any 
change in the power conservation equation. 

\subsection{Polarization States and Duality transformation}

Following \cite{cf} and rearranging terms of the components of the wave 
equations (27) and (28) (following the procedure in \cite{cf}), we obtain  
the following equations for the polarized states
\begin{eqnarray}
\ddot{b}_{\pm} ~\mp~ 2 i H' \dot{b}_{\pm} ~\mp~ 2 i \dot{H}' b_{\pm} ~&=&~ ~b_{\pm}''
~\pm~ 2 i \dot{H} b_{\pm}' ~+~ 2 H'' e_{\pm}  \\  
\ddot{e}_{\pm} ~\mp~ 2 i H' \dot{e}_{\pm} ~\mp~ 2 i \dot{H}' e_{\pm} ~&=&~ ~e_{\pm}''
~\pm~ 2 i \dot{H} e_{\pm}' ~-~ 2 \ddot{H} b_{\pm},    
\end{eqnarray}
where 
\begin{eqnarray}
b_{\pm}(\eta,z) &=& \tilde{B}_{x}(\eta,z) ~\pm~ i \tilde{B}_{y}(\eta,z) \nonumber\\
e_{\pm}(\eta,z) &=& \tilde{E}_{x}(\eta,z) ~\pm~ i \tilde{E}_{y}(\eta,z).
\end{eqnarray}

Note that the Eqs.(35) and (36) are converted into each other by the
transformation ~$e_{\pm} \rightarrow b_{\pm}$, $b_{\pm} \rightarrow
- e_{\pm}$~ provided the equation ~$\Box H ~\equiv~ {\ddot H} ~-~ H'' 
~=~0$~ is obeyed. This is the usual electro-magnetic {\it duality} symmetry 
of the Maxwell equations. The Maxwell-KR system indeed possess this 
invariance in a flat spacetime background with cosmplogical scale factor
~$R ~=~ 1$, ~as is evident from Eq.(20). In a curved spacetime
background, however, one does not have this invariance any more.

Rewriting the Eqs.(35) and (36) as follows:
\begin{eqnarray}
\ddot{b}_{\pm} ~\mp~ 2 i H' \dot{b}_{\pm} ~\mp~ 2 i \dot{H}' b_{\pm} ~&=&~ ~b_{\pm}''
~\pm~ 2 i \dot{H} b_{\pm}' ~+~ 2 H'' e_{\pm} ~=~ \alpha_{\pm}(\eta,z) \\  
\ddot{e}_{\pm} ~\mp~ 2 i H' \dot{e}_{\pm} ~\mp~ 2 i \dot{H}' e_{\pm} ~&=&~ ~e_{\pm}''
~\pm~ 2 i \dot{H} e_{\pm}' ~-~ 2 \ddot{H} b_{\pm} ~=~ \beta_{\pm}(\eta,z),   
\end{eqnarray}
We seek the appropriate forms of the functions $\alpha(\eta,z)$ and
$\beta(\eta,z)$ by examining the limiting forms of the above equations:

{\bf I.~ Limit $H' \rightarrow 0$ : \/}

In this limit, where $H$ becomes purely a function of $\eta$, the equations (37)
and (38) reduce to
\begin{eqnarray}
\ddot{b}_{\pm}  &=& ~b_{\pm}'' ~\pm~ 2 i \dot{H} b_{\pm}' ~=~ 
\alpha_{\pm}(H' \rightarrow 0) \\  
\ddot{e}_{\pm}  &=& ~e_{\pm}'' ~\pm~ 2 i \dot{H} e_{\pm}' ~-~ 2 \ddot{H} b_{\pm} ~=~ 
\beta_{\pm}(H' \rightarrow 0)   
\end{eqnarray}

Assuming solutions of the form
\begin{eqnarray}
b_{\pm}(\eta,z) &=& ~b_{0}^{\pm}(\eta)~ e^{- i k z} \\
e_{\pm}(\eta,z) &=& ~e_{0}^{\pm}(\eta)~ e^{- i k z}
\end{eqnarray}
we find
\begin{eqnarray}
\ddot{b}_{\pm}  &=& ~ - (k^{2} \mp 2 k \dot{H}) b_{\pm} ~=~ 
\alpha_{\pm}(H' \rightarrow 0) \\  
\ddot{e}_{\pm}  &=& ~ - (k^{2} \mp 2 k \dot{H}) e_{\pm} ~-~ 2 \ddot{H} b_{\pm} ~=~ 
\beta_{\pm}(H' \rightarrow 0).   
\end{eqnarray}

{\bf II.~ Limit $\dot{H} \rightarrow 0$ : \/}

This limiting case implies $H$ to be a function of $z$ only but as is evident
from Eq.(22) $H'$ is merely a constant. Therefore Eqs.(37) and (38) reduce to
\begin{eqnarray}
\ddot{b}_{\pm} ~\mp~ 2 i H' \dot{b}_{\pm} &=& ~b_{\pm}'' ~=~ 
\alpha_{\pm}(\dot{H} \rightarrow 0) \\  
\ddot{e}_{\pm} ~\mp~ 2 i H' \dot{e}_{\pm} &=& ~e_{\pm}'' ~=~ 
\beta_{\pm}(\dot{H} \rightarrow 0)
\end{eqnarray}

Again assuming solutions of the form
\begin{eqnarray}
b_{\pm}(\eta,z) &=& ~b_{1}^{\pm}(z)~ e^{ i k \eta} \\
e_{\pm}(\eta,z) &=& ~e_{1}^{\pm}(z)~ e^{ i k \eta}
\end{eqnarray}
we get
\begin{eqnarray}
b_{\pm}''  &=& ~ - (k^{2} \mp 2 k H') b_{\pm} ~=~ 
\alpha_{\pm}(\dot{H} \rightarrow 0) \\  
e_{\pm}''  &=& ~ - (k^{2} \mp 2 k H') e_{\pm} ~=~ 
\beta_{\pm}(\dot{H} \rightarrow 0).   
\end{eqnarray}

By looking at these limiting forms of $\alpha_{\pm}$ and $\beta_{\pm}$ 
given in Eqs.(44),(45) and in Eqs.(50),(51) it seems reasonable to suggest the
following simplest possible structures of these functions:
\begin{eqnarray}
\alpha_{\pm}(\eta,z) &=& ~- \left[ k^{2} \mp 2 k ( \dot{H} + H' ) \right]
b_{\pm}(\eta,z) \\
\beta_{\pm}(\eta,z) &=& ~- \left[ k^{2} \mp 2 k ( \dot{H} + H' ) \right]
e_{\pm}(\eta,z) ~-~ 2 \ddot{H} b_{\pm}(\eta,z). \\
\end{eqnarray}
Moreover, setting
\begin{equation}
e_{\pm} (\eta,z) = a_{\pm} (\eta,z)~ b_{\pm} (\eta,z)
\end{equation}
we write the equations (37) and (38) in a more elegant form
\begin{eqnarray}
\ddot{b}_{\pm} ~\mp~ 2 i H' \dot{b}_{\pm} ~+~ \left[ k^{2} ~\mp~ 2 k 
( \dot{H} + H' ) ~\mp~ 2 i \dot{H}' \right] b_{\pm}(\eta,z) &=& 0 \\  
b_{\pm}'' ~\pm~ 2 i \dot{H} b_{\pm}' ~+~ \left[ k^{2} ~\mp~ 2 k 
( \dot{H} + H' ) ~+~ 2 H'' a_{\pm} \right] b_{\pm}(\eta,z) &=& 0;
\end{eqnarray}
and
\begin{eqnarray}
\ddot{e}_{\pm} ~\mp~ 2 i H' \dot{e}_{\pm} ~+~ \left[ k^{2} ~\mp~ 2 k 
( \dot{H} + H' ) ~\mp~ 2 i \dot{H}' ~+~ 2 \frac{\ddot{H}}{a_{\pm}} \right] 
e_{\pm}(\eta,z) &=& 0 \\  
e_{\pm}'' ~\pm~ 2 i \dot{H} e_{\pm}' ~+~ \left[ k^{2} ~\mp~ 2 k 
( \dot{H} + H' ) \right] e_{\pm}(\eta,z) &=& 0.
\end{eqnarray}

\section{Flat Spacetime Background}

In order to get a preliminary idea as to how the coupling of a spacetime 
dependent KR field to Einstein-Maxwell theory affects the electromagnetic 
waves, and thereby effects in an optical activity in the radiation coming 
from distant galactic sources, we consider the simplest situation --- that 
is, of a flat universe with cosmological scale factor $R(\eta) =
1$. Admittedly, quantitative details of results of this section are
cosmologically untenable for obvious reasons.

The equation of motion (22) for $H$ can be solved readily to obtain
\begin{equation}
H(\eta,z) ~=~ (c_{1} \sin p \eta ~+~ c_{2} \cos p \eta)~ \cos p z
\end{equation}
where $c_{1}$ and $c_{2}$ are arbitrary integration constants. Demanding that 
the above solution must reduce to the form $( h \eta ~+~ h_{0} )$
which is the solution of Eq.(22) in the limit $p \rightarrow 0$  we infer
$c_{1} = h/p$ and $c_{2} = h_{0}$. Here $h$ and $h_{0}$ are the same arbitrary
constants denoted in \cite{skpm}. Making a Taylor series expansion of the 
various functions appearing in Eq.(60) around $p = 0$ we write
\begin{equation}
H(\eta,z) ~=~ (h \eta ~+~ h_{0}) ~-~ \frac{p^{2}}{2} \left(\frac{h \eta^{3}}{3}
~+~ h_{0} \eta^{2} + h \eta z^{2} + h_{0} z^{2}\right) ~+~ O(p^{4})
\end{equation} 
  
Substituting this $H$ in Eqs.(55) and (57) and assuming solution of the standard
WKB type given by
\begin{equation}
b_{\pm} (\eta,z) ~=~ \bar{b} ~ e^{i k~ S_{\pm} (\eta,z)}
\end{equation}
with
\begin{equation}
S_{\pm} ~=~ S_{0}^{\pm} ~+~ \frac{S_{1}^{\pm}}{k} ~+~ \frac{S_{2}^{\pm}}{k^2} 
~+~ \cdots 
\end{equation}

\noindent
we get after partial integrations of Eqs.(56) and (57) with respect to 
$\eta$ and $z$ respectively
\begin{equation}
b_{\pm} (\eta,z) ~=~ \bar{b}~ \exp \left\{ i k \eta ~\mp~ i \left[ h \eta
~-~
\frac{ p^2}{2} \left( \frac{ h \eta^3}{3} ~+~ h_{0} \eta^2 ~+~ h z^2 \eta \right) 
~+~ O(p^4) \right] ~+~ O\left(\frac 1 k \right) ~+~ i k f_{\pm}
(z) \right\}
\end{equation}
and 
\begin{equation}
b_{\pm} (\eta,z) ~=~ \bar{b}~ \exp \left \{ i k g_{\pm} (\eta) ~-~ i k z
~\mp~ 
i \left[ \frac{ p^2}{2} ( h \eta ~+~ h_{0} ) z^2 ) ~+~ O(p^4) \right] ~+~ 
O\left(\frac 1 k \right)  \right \}.
\end{equation}

Comparing these two expressions we assert the forms of the arbitrary functions
$f_{\pm} (z)$ and $g_{\pm} (\eta)$ and write $b_{\pm} (\eta,z)$ as follows:
\begin{equation}
b_{\pm} (\eta,z) ~=~ \bar{b}~ \exp \left \{ i k ( \eta - z ) ~\mp~ 
i \left[ h \eta ~-~ \frac{ p^2}{2} \left( \frac{ h \eta^3}{3} ~+~ h_{0} \eta^2 
~-~ h_{0} z^2 \right) ~+~ O(p^4) \right] ~+~ O\left(\frac 1 k
\right) \right \}
\end{equation}
 
A similar approach for $e_{\pm} (\eta,z)$ yields
\begin{equation}
e_{\pm} (\eta,z) ~=~ \bar{e}~ \exp \left \{ i k ( \eta - z ) ~\mp~ 
i \left[ h \eta ~-~ \frac{ p^2}{2} \left( \frac{ h \eta^3}{3} ~+~ h_{0} \eta^2 
~-~ h_{0} z^2 \right) ~+~ O(p^4) \right] ~+~ O\left(\frac 1 k
\right) \right \}
\end{equation}
which differs from Eq.(66) only in the constant coefficient $\bar{e}$ 
and in the higher order $O\left(\frac 1 k \right)$. However, it should be 
mentioned here that while using the WKB technique we are assuming that the
function $a_{\pm}$ is not increasing rapidly as $k$ increases. In fact, in
absence of the KR field, when we have the plane wave solutions of Maxwell's 
equations, $a_{+} ~=~ a_{-} ~=~ \bar{e} / \bar{b} ~=~$ constant. The spacetime
dependence of $a_{\pm}$ comes only in presence of a spacetime-dependent $H$. 
Since $H$ is being treated as a small purturbation over the Maxwell field, it
is rather plausible to think $a_{\pm}$ to be not much different from the 
constant $\bar{e} / \bar{b}$ and a very slowly-varying function of spacetime.
But the constant $\bar{e} / \bar{b}$ is arbitrary and cannot generically
be argued as increasing with $k$. Therefore, the assumption that $a_{\pm}$
remains quite invariant as $k$ increases is fairly justified. Using WKB method,
the solutions obtained above involves $a_{\pm}$ only in the higher order
$O\left(\frac 1 k \right)$.

Now, the circular polarization states are defined by $b_{\pm}$ and $e_{\pm}$ and 
the extent of the optical birefringence due the presence of the KR field can be 
estimated directly by calculating the rotation angle of the plane of polarization 
of the electromagnetic wave, which is given by the phase difference ~ $\phi_{mag}~ 
\equiv~ \frac 1 2 [arg ~b_{+} - arg ~b_{-}] $ for the magnetic field and $\phi_{elec}~ 
\equiv~ \frac 1 2 [arg ~e_{+} - arg ~e_{-}] $ for the electric field. The phase shift 
is given by
\begin{equation}
\phi_{mag} (\eta,z) ~\approx~ \phi_{elec} (\eta,z) ~\approx~ - h \eta ~+~ \frac{p^2}{2}
\left( \frac{h \eta^3}{3} ~+~ h_{0} \eta^2 ~-~ h_{0} z^2 \right) 
\end{equation}
for $~ h,p << k$.

It is interesting to see that the change in the rotation angle, calculated
here, over that found in \cite{skpm} for flat universe, is primarily given 
by the $p^2$-dependent part. But as $p$ is considered to be very small we
infer that this change is rather insignificant.

\section{Spatially Flat FRW Spacetime Background}

We now turn to less trivial background spacetimes. We consider a spatially
flat expanding universe dominated by radiation and matter, in turn.

\subsection{Radiation dominated Universe}

The scale factor for this model, in real time, is given by 
\begin{equation}
~[R(\eta)]^{RD} ~=~ \frac{\eta}{\eta_{r}}  
\end{equation}
where $\eta_{r} = \left(8 \pi G \epsilon_{0} /3\right)^{-1/3}$, $\epsilon_{0}$ 
being the primordial radiant energy density.

Substituting this in the equation of motion (22) for $H$ we obtain 
\begin{equation}
\eta^{2} \ddot{H} + 2\eta \dot{H} + p^{2} \eta^{2} H ~=~ 0
\end{equation}
which has the form of a transformed Bessel equation with solution
\begin{equation}
H (\eta,z) ~=~ \eta^{- \frac 1 2}~\left[ \bar{c}_{1} ~ J_{1/2} (p \eta)
~+~ \bar{c}_{2} ~Y_{1/2} (p \eta) \right].
\end{equation}
    
Simplifying the Bessel functions of first and second kinds, viz., $J$ and $Y$,
the above solution can be written as
\begin{equation}
H (\eta,z) ~=~ \frac 1 \eta ~( c_{1} ~\sin ~p \eta ~+~ c_{2} ~ \cos ~p \eta )
~ \cos~p z.
\end{equation}
Imposing again the boundary condition that this must reduce to the limiting
form $ ~\left[ -~ \frac {h \eta_{r}^2}{\eta} ~+~ h_{0} \right]~$ \----
the solution 
of Eq.(22) \---- ~as $~p \rightarrow 0~$, we set $~c_{1} ~=~ \frac
{h_{0}} p~$ and 
$~c_{2} ~=~ - h_{r}$, ~with $~h_{r} ~=~ h~\eta_{r}^2$, ~whence
\begin{equation}
H (\eta,z) ~=~ \frac 1 \eta ~\left( \frac {h_{0}} p ~\sin ~p \eta ~-~ h_{r} 
~ \cos ~p \eta \right)~ \cos~p z.
\end{equation}

Plugging in the Taylor expanded form of this $H$ in Eqs.(56) - (59) and using the 
same WKB technique as for the flat universe, we obtain
\begin{equation}
b_{\pm} (\eta,z) ~=~ \bar{b}~ \exp \left[ i k ( \eta - z ) ~\pm~ 
i \left[ \frac {h_{r}} \eta ~+~ \frac{p^{2}}{2} \left( \frac{h_{0} \eta^2}{3} 
~-~ h_{r} \eta ~-~ h_{0} z^2 \right) ~+~ O(p^4) \right] ~+~ O\left(\frac 1 k \right) 
\right]
\end{equation}
and similar expression for $e_{\pm} (\eta,z)$.

The phase shift in this case is given by
\begin{equation}
\phi_{mag} (\eta,z) ~\approx~ \phi_{elec} (\eta,z) ~\approx~ \frac {h_{r}} \eta 
~-~ \frac{p^2}{2} \left( h_{r} \eta ~-~ \frac{h_{0} \eta^2}{3} ~+~ h_{0} z^2 \right) 
\end{equation}
for $ ~h,p ~<<~ k$ and $~h_{r} ~=~ h~ \eta_{r}^2$.

\subsection{Matter dominated Universe}

In this case, where the scale factor is governed by
\begin{equation}
~[R(\eta)]^{MD} ~=~ \frac{\eta^2}{\eta_{m}^2}  
\end{equation}
with $\eta_{m} = \left(8 \pi G \rho_{0} /3\right)^{-1/3}$ ~($\rho_{0}$ 
~ \--- the initial matter density), ~Eq.(22) reduces again to a transformed 
Bessel equation
\begin{equation}
\eta^{2} \ddot{H} + 4\eta \dot{H} + p^{2} \eta^{2} H ~=~ 0
\end{equation}
having simplified solution
\begin{equation}
H (\eta,z) ~=~ \frac 1 {\eta^3} ~[ (c_{1} ~-~ c_{2}~p \eta) ~\sin ~p \eta ~-~
(c_{2} ~+~ c_{1}~p \eta) ~ \cos ~p \eta ]~ \cos~p z.
\end{equation}

Determining the constants $c_{1}$ and $c_{2}$ using, as before, the
boundary 
condition on $H$ in the limit $~p \rightarrow 0~$, we write
\begin{equation}
H (\eta,z) ~=~ - ~\frac {h_{m}} {3 \eta^3} ~( \cos ~p \eta ~+~ p \eta ~\sin ~
p \eta)~ \cos~p z.
\end{equation}
\noindent
where $~h_{m} ~=~ h \eta^4$. With this $H$, we obtain using the WKB method
\begin{equation}
b_{\pm} (\eta,z) ~=~ \bar{b}~ \exp \left[ i k ( \eta - z ) ~\pm~ 
i \left[ \frac {h_{m}} {3 \eta^3} ~+~ p^{2} h_{m} \left( \frac 1 \eta 
~+~ \frac {z^2} 6 \right) ~+~ O(p^4) \right] ~+~ O\left(\frac 1 k \right) 
\right]
\end{equation}
and similar expression for $e_{\pm} (\eta,z)$. The phase shift can be calculated

\begin{equation}
\phi_{mag} (\eta,z) ~\approx~ \phi_{elec} (\eta,z) ~\approx~ \frac {h_{m}}{3 \eta^3} 
~+~ p^2 h_{m} \left( \frac 1 \eta ~+~ \frac {z^2} 6 \right) 
\end{equation}

\noindent
for $ ~h,p ~<<~ k$.

\section{Conclusions}

One rather surprising aspect of our finding is the loss of orthogonality of the
electric and magnetic vectors in the radiation field which exists even in the
case of a spatially homogeneous axion field, so long as the axion is
time-dependent. This will indeed affect the measurement of the precise rotation
of the plane of polarization, although for large redshift sources to which we
confine, this effect may be ignored for all practical purposes. 

In general the electric and magnetic vectors seem to have
solutions generically represented as 
\begin{equation}
(Electric/Magnetic ~field ~combinations) ~=~ 
(constant) ~e^{i(kz ~-~ \omega \eta)}~e^{i H(\eta, z)}
\end{equation}

The additional contribution to the angle of rotation arising out of the spatial
inhomogeneity introduced in this paper is actually quite small under our
assumption that the wavelength of the axion wave is far larger than the
radiation. Thus from an observational standpoint this new effect is not too
significant, although it is still quite distinct in its behaviour from Faraday
rotation.

We have used standard WKB type methods to arrive at some solution to
the complicated set of equations which arise even at the lowest order. Among 
other things we have explored the properties of the electric and magnetic
fields of the Maxwell--KR system. We also checked the validity of the
standard electric--magnetic duality and pointed out the existence of a
very simple solution to the  complicated equations.

\section{Acknowledgements}

This work is supported by Project grant no. 98/37/16/BRNS cell/676 from The 
Board of Research in Nuclear Sciences, Department of Atomic Energy, Government
of India, and the Council of Scientific and Industrial Research, Government of 
India.

\end{document}